\documentclass[pre,aps,reprint,amsmath]{revtex4-1}
\usepackage[T1]{fontenc} 
\usepackage{graphicx}
\usepackage[utf8]{inputenc}
\usepackage{lmodern}
\usepackage{bm}
\usepackage{bbold}
\usepackage[hidelinks]{hyperref}				
\usepackage{color}
\usepackage{ulem}

\newcommand{\rv}{{\bf r}}

\newcommand{\chimu}{\chi_\mu}
\newcommand{\chiT}{\chi_T}
\newcommand{\rhostar}{\chi_\star}
\newcommand{\chiTb}{\chi_T^b}

\renewcommand{\alpha}{a}

\newcommand{\cov}{{\,\rm cov}}

\begin{document}
\title{Fluctuation profiles in inhomogeneous fluids}

\author{Tobias Eckert$^*$}
\affiliation{Theoretische Physik II, Physikalisches Institut, 
  Universit{\"a}t Bayreuth, D-95447 Bayreuth, Germany}

\author{Nico C. X. Stuhlm\"uller$^*$}
\affiliation{Theoretische Physik II, Physikalisches Institut,
  Universit{\"a}t Bayreuth, D-95447 Bayreuth, Germany}

\author{Florian Samm\"uller$^*$}
\affiliation{Theoretische Physik II, Physikalisches Institut,
  Universit{\"a}t Bayreuth, D-95447 Bayreuth, Germany}

\author{Matthias Schmidt}
\affiliation{Theoretische Physik II, Physikalisches Institut, 
  Universit{\"a}t Bayreuth, D-95447 Bayreuth, Germany}
\email{Matthias.Schmidt@uni-bayreuth.de}

\date{30 September 2020,  revised version: 23 November 2020, to appear
in Physical Review Letters}

\begin{abstract}
Three one-body profiles that correspond to local fluctuations in
energy, in entropy, and in particle number are used to describe the
equilibrium properties of inhomogeneous classical many-body
systems. Local fluctuations are obtained from thermodynamic
differentiation of the density profile or equivalently from average
microscopic covariances. The fluctuation profiles follow from
functional generators and they satisfy Ornstein-Zernike relations.
Computer simulations reveal markedly different fluctuations in
confined fluids with Lennard-Jones, hard sphere, and Gaussian core
interactions.
\end{abstract}

\maketitle

Inhomogeneous fluids comprise a large class of relevant and
fundamental physical systems, in which a broad range of phenomena and
underlying mechanisms occur \cite{hansen2013,evans1979,
  evans2016specialIssue}. Examples include the behaviour of fluids in
narrow confinement \cite{nygard2016prx}, electrolytes near surfaces
\cite{martinjimenez2017natCom}, dense fluid structuring as revealed in
atomic force microscopy \cite{hernandez-munoz2019}, thermal resistance
of liquid-vapor interfaces \cite{muscatello2017}, nonequilibrium
steady states in active \cite{paliwal2018,rodenburg2018}, sheared
\cite{brader2011shear} and driven \cite{scacchi2018} fluids, as well
as the orientation-resolved ordering of water around complex solutes
\cite{jeanmairet2013jcp,jeanmairet2013jpcl,levesque2012jcp,
  sergiievskyi2014,jeanmairet2019capacitor}.  The average one-body
density distribution, or short the density profile, is used as the
standard tool for analyzing such inhomogeneous systems.

In particular the occurrence of hydrophobicity
\cite{levesque2012jcp,jeanmairet2013jcp,stewart2012pre,
  stewart2014jcp,evans2015jpcm,evans2015prl,
  chacko2017,evans2017,evans2019pnas,rensing2019pnas,evans2016prl,
  giacomello2016,giacomello2019}, and its important consequences in
biological systems, have been at the center of much current scientific
attention and debate. At hydrophobic substrates or around hydrophobic
solutes, water avoids contact of its liquid phase. In the more general
framework of solvophobicity (where the liquid is not necessarily
water), Evans and coworkers argue
\cite{stewart2012pre,stewart2014jcp,evans2015jpcm,evans2015prl,
  chacko2017,evans2017,evans2019pnas,rensing2019pnas,evans2016prl}
that the {\it local} compressibility is a more suitable indicator of
the occurrence of drying than is the bare density profile. These
authors obtain (and define) the local compressibility $\chimu(\rv)$ by
differentiating the density profile, $\rho(\rv)$, with respect to the
chemical potential, $\mu$, in straightforward generalization of the
definition of the bulk compressibility; they also use correlators and
reweighting to obtain $\chi_\mu(\rv)$.  At a planar substrate
$\chimu(\rv)$ measures in-plane density fluctuations and the results
show a very pronounced signal when a drying film develops near the
substrate. The findings of
Refs.~\cite{stewart2012pre,stewart2014jcp,evans2015jpcm,evans2015prl,
  chacko2017,evans2017,evans2019pnas,rensing2019pnas,evans2016prl},
obtained over a range of microscopic models (differently truncated
Lennard-Jones particles, as well as classical models for water)
convincingly demonstrate the superiority of $\chimu(\rv)$ over
$\rho(\rv)$ as an indicator function for the occurring physics.

From a fundamental point of view, and in particular that of classical
density functional theory
\cite{evans1979,hansen2013,evans2016specialIssue} (DFT) where
$\rho(\rv)$ is the central (variational) variable, the above situation
is perplexing, and it is unclear whether the observation is merely
relevant for the particular situations they consider or whether it is
indicative of a more general underlying theoretical structure.

Here we show that the latter is the case, when the local
compressibility $\chimu(\rv)$ is complemented by two further local
measures of fluctuations. One of these additional fields is the local
thermal susceptibility $\chiT(\rv)$, which constitutes the partial
derivative with respect to temperature $T$ of the density profile. As
we demonstrate below, $\chiT(\rv)$ is indicative of entropic
correlation effects. It is then natural to also consider a reduced
density profile $\rhostar(\rv)$, where the thermal and chemical
fluctuations have been subtracted. Hence
\begin{align}
  \chi_\mu(\rv) &= \frac{\partial \rho(\rv)}{\partial\mu}\Big|_T,
  \label{EQparametricDerivative1}\\
  \chi_T(\rv) &= \frac{\partial \rho(\rv)}{\partial T}\Big|_\mu,
  \label{EQparametricDerivative2}\\
  \rhostar(\rv) &=   \rho(\rv)
  -\mu \frac{\partial \rho(\rv)}{\partial \mu}\Big|_T
  -T \frac{\partial \rho(\rv)}{\partial T}\Big|_\mu
  \label{EQparametricDerivative3}\\
  &\equiv \rho(\rv)-\mu \chi_\mu(\rv)-T \chi_T(\rv),
  \label{EQparametricDerivative4}
\end{align}
where the external potential $V_{\rm ext}(\rv)$ is kept constant under
the partial thermodynamic derivatives. Equation
\eqref{EQparametricDerivative3} is akin to a Legendre transform of the
density profile with respect to the thermodynamic variables $T$ and
$\mu$, and \eqref{EQparametricDerivative4} is obtained from
\eqref{EQparametricDerivative3} by using
\eqref{EQparametricDerivative1} and \eqref{EQparametricDerivative2}.

Given the relationships of the three {\it fluctuation profiles}
$\chi_\alpha(\rv)$, $\alpha=\mu,T,\star$ to the density profile
\eqref{EQparametricDerivative1}--\eqref{EQparametricDerivative3}, we
demonstrate three further fundamental properties: i)~Representation as
explicit correlation functions, given as ensemble averages, which
makes all three correlators directly accessible in particle-based
simulations via averaging. ii) All three fluctuation profiles can be
generated as response functions to changes in the external potential
$V_{\rm ext}(\rv)$. iii) The fluctuation profiles satisfy
Ornstein-Zernike (OZ) relations, which remarkably have simpler
structure than the standard (inhomogeneous) OZ relation
\cite{evans1979,hansen2013,ornstein1914}. We demonstrate, based on
computer simulation data, that the fluctuation profiles are highly
sensitive to the type of interparticle interactions, and that they
display markedly different behaviour for different model fluids.

Recall that the density profile $\rho(\rv)$ measures the
microscopically resolved mean number of particles at position~$\rv$.
Its integral over the system volume $V$ yields the average total
number of particles, $\bar N=\int_V d\rv \rho(\rv)$, and for bulk
fluids $\rho(\rv)=\rho_b=\rm const$, where $\rho_b=\bar N/V$ is the
bulk fluid (number) density. In the grand ensemble, when the system is
coupled to a heat bath at temperature $T$ and to a particle bath at
chemical potential $\mu$, then $\rho_b(T,\mu)$ represents a
fundamental equation of state of the bulk liquid, from which all
further thermodynamic quantities can be obtained. Differentiation of
$\rho_b$ with respect to the thermodynamic variables yields the bulk
thermal susceptibility when changing temperature, $\chiTb = \partial
\rho_b/\partial T|_\mu$, and the isothermal compressibility when
changing the chemical potential, $\chimu^b = \partial \rho_b/\partial
\mu|_T$. Hence $\chimu^b$ has the status of a chemical
susceptibility. The respective right hand sides imply that $\chiTb$
and $\chimu^b$ are global response functions that characterize the
ease (or lack thereof) to influence the bulk density upon changing the
control parameter of either bath.

To be specific, we consider Hamiltonians of the form $\hat H=\hat
K+u(\rv^N)+\sum_i V_{\rm ext}(\rv_i)$, where $\hat K$ indicates
kinetic energy and $u(\rv^N)$ denotes the interparticle interaction
potential that depends on all $N$ particle positions
$\rv^N\equiv\rv_1\ldots\rv_N$, where $\rv_i$ is the position of
particle $i=1\ldots N$ in $D$ spatial dimensions. In order to resolve
the density locally it is common to introduce the density {\it
  operator} $\hat\rho(\rv)=\sum_i\delta(\rv-\rv_i)$. The density {\it
  profile} is then obtained as the average
$\rho(\rv)=\langle\hat\rho(\rv)\rangle$, where the angular brackets
denote an average over microstates that are distributed according to
the (grand canonical) equilibrium distribution function
$\Psi=\exp(-\beta(\hat H-\mu N))/\Xi$, where $\beta=1/(k_BT)$, with
$k_B$ indicating the Boltzmann constant and $\Xi$ the grand canonical
partition sum.

We start by considering correlators, i.e.\ ensemble averages over
suitable many-body (phase space) functions. Differentiating the
density profile in the form $\rho(\rv)=\langle\hat\rho(\rv)\rangle$
with respect to the thermodynamic parameters according to
\eqref{EQparametricDerivative1} and \eqref{EQparametricDerivative2}
naturally leads to results that are of covariance form. We indicate
the covariance of two operators (phase space functions) $\hat A$ and
$\hat B$ as $\cov(\hat A,\hat B)=\langle\hat A\hat B\rangle-\langle
\hat A \rangle \langle \hat B \rangle$.  We obtain
\begin{align}
  \chimu(\rv)  &= \beta\cov(N,\hat\rho),
  \label{EQchimuAsCorrelator}
  \\
  \chiT(\rv) &= \beta \cov(\hat S,\hat\rho)
  \label{EQchiTAsCorrelatorEntropicForm}
  \\ 
  &\equiv \beta\cov(\hat H - \mu N,\hat\rho)/T,
  \label{EQchiTAsCorrelatorEnergyForm}  
  \\
  \rhostar(\rv) &=  \rho(\rv) 
  -\beta \cov(\hat H,\hat\rho),
  \label{EQrhoStarAsCorrelator}
\end{align}
where $\hat S = -k_B \ln \Psi$ is the entropy operator, such that
$S=\langle\hat S\rangle$ is the total entropy.  The form
\eqref{EQchimuAsCorrelator} has been given before in
Refs.~\cite{evans2015prl,evans2017}, while
\eqref{EQchiTAsCorrelatorEnergyForm} is obtained by inserting the
explicit Boltzmann form of $\Psi$ into
\eqref{EQchiTAsCorrelatorEntropicForm}. Crucially
\eqref{EQchiTAsCorrelatorEnergyForm} can be carried out via importance
sampling in particle-based simulations (which is hampered in
\eqref{EQchiTAsCorrelatorEntropicForm} due to the poor direct
accessibility of $\ln\Psi$).  Note that
\eqref{EQchiTAsCorrelatorEntropicForm} is different from the entropy
density $\langle\hat \rho\hat S/N\rangle$ of
Ref.~\cite{schmidt2011pre}.  It becomes apparent that
\eqref{EQparametricDerivative3} leads to
\eqref{EQrhoStarAsCorrelator}, which constitutes a reduced density
distribution, where the energy-density covariance is subtracted from
the full density profile.

\begin{figure*}[bt]
  \includegraphics[width=0.9\textwidth]{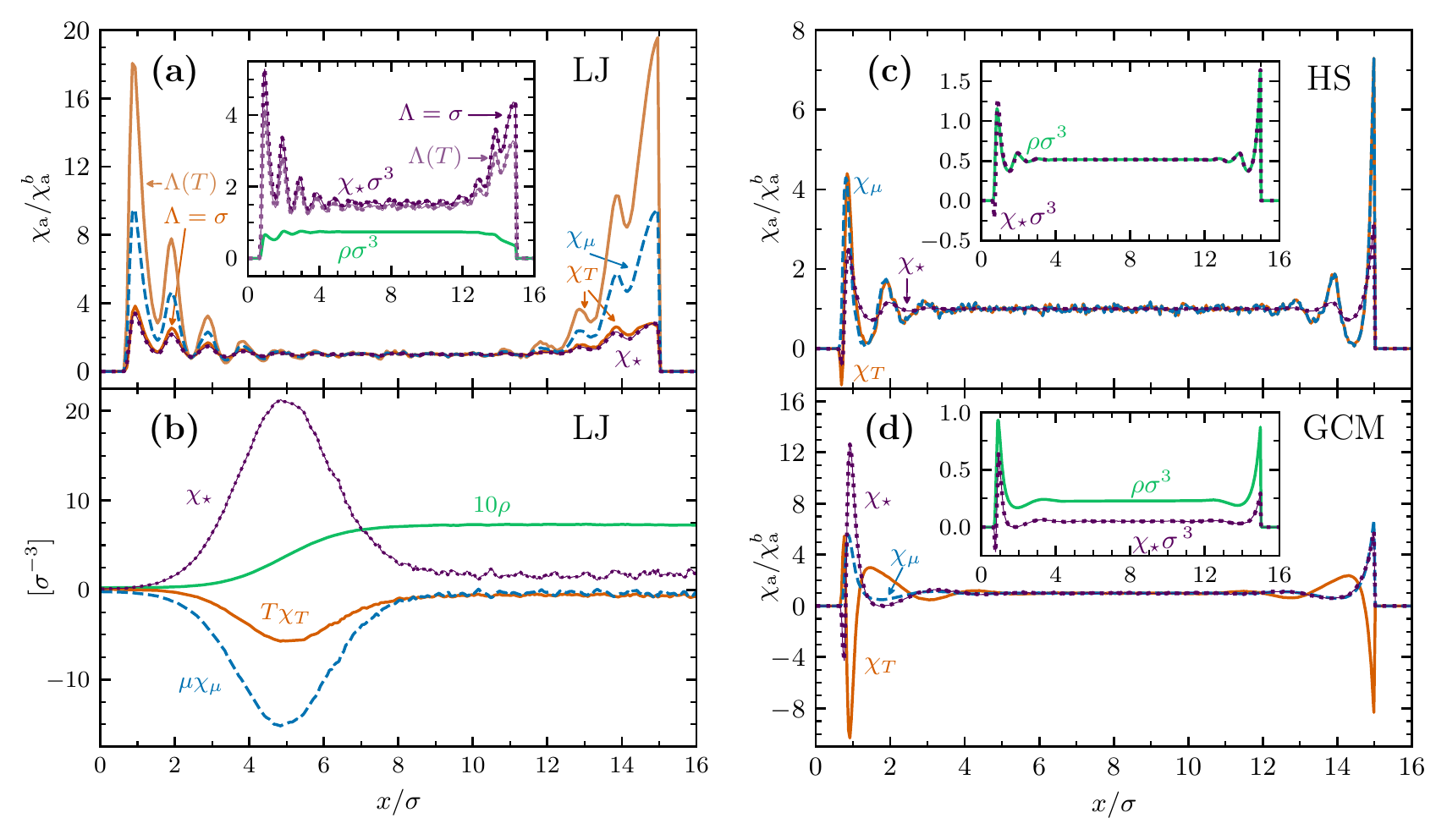}
  \caption{Normalized fluctuation profiles
    $\chi_\alpha(x)/\chi_\alpha^b$, where $\alpha=\mu,T,\star$, for
    the LJ liquid (a,b), the HS fluid (c), and the GCM (d). Results
    are for fluids confined between a planar 3-9 LJ wall (left) and a
    planar hard wall (right) (a,c,d) and for the LJ free gas-liquid
    interface (b).  Shown are results for $\chiT$ (orange solid line),
    $\chimu$ (blue dashed line) and $\rhostar$ (purple dotted line) as
    a function of distance $x/\sigma$ across the slit and normalized
    by the respective bulk value. The insets in (a,c,d) show the
    corresponding density profile $\rho(x)\sigma^3$ (solid green line)
    and replot the (non-normalized) reduced density profiles
    $\rhostar(x)\sigma^3$ using either convention for $\Lambda$ in (a)
    (as indicated); the results in (b,c,d) are for $\Lambda=\sigma$;
    The bulk values are $\chi_T^b \epsilon\sigma^3/k_B = -0.640$ (a),
    $-0.217$ (c), 0.082 (d); $\chi_\mu^b \epsilon\sigma^3 = 0.092$
    (a), 0.065 (c), 0.063 (d), and $\rhostar^b \sigma^3= 1.553$ (a),
    0.515 (c), 0.051 (d) for $\Lambda=\sigma$; for $\Lambda(T)$ in
    (a): $\chi_T^b \epsilon\sigma^3/k_B = 0.432$ and $\rhostar^b
    \sigma^3= 1.435$.}
  \label{figLJ}
\end{figure*}

For the ideal gas ($u\equiv 0$) one can show $\chimu^{\rm
  id}(\rv)=\beta\rho(\rv)$ and also obtain closed forms for
$\chiT(\rv)$ and $\rhostar(\rv)$ \cite{SMchilocal}.  Turning to
interacting systems, we first consider the Lennard-Jones (LJ) liquid
confined in an asymmetric planar slit pore, which consists of two
opposing walls, inspired by Ref.~\cite{evans2015jpcm}.  We use grand
canonical Monte Carlo simulations in $D=3$ dimensions, with simulation
box dimensions $L_x = 15\sigma$ and $L_y = L_z = 6\sigma$ and periodic
boundaries in the $y$- and $z$-directions.  Here $\sigma$ is the LJ
length scale. We equilibrate for $\geq 10^7$ grandcanonical MC moves
before sampling data $10^8-10^9$ times, with $100-300$ trial moves
between consecutive samples.  We consider a 3-9 LJ wall at $x=0$,
described by an external potential $V_{\rm ext}^{\rm
  LJ}(x)=(\epsilon_w/2)[(\zeta/x)^9-3(\zeta/x)^3]$, where the range
$\zeta$ is set to $\zeta=0.95\sigma$ \cite{evans2015jpcm}.  The
strength is chosen as $\epsilon_w=0.49\epsilon$, where $\epsilon$ is
the LJ energy scale. This value of $\epsilon_w$ corresponds to a
weakly attractive wall; it is intermediate between the solvophobic and
neutral cases of Ref.~\cite{evans2015jpcm}.  Both $V_{\rm ext}^{\rm
  LJ}(x)$ and the LJ pair potential are cut and shifted
\cite{hansen2013} at a cutoff distance $2.5\sigma$. The second wall is
hard and located at $x=L_x$, such that no particle center can go
beyond that distance. We choose a statepoint on the liquid side of the
gas-liquid binodal: $k_BT = 0.85\epsilon$ and $\tilde\mu=
-2.90\epsilon$; here we use a reduced chemical potential with the
kinetic contribution subtracted, $\tilde\mu\equiv \mu - k_BT D
\ln(\Lambda/\sigma)$, where $\Lambda$ indicates the thermal de Broglie
wavelength.

Fig.~\ref{figLJ}(a) illustrates the resulting behaviour of the three
fluctuation profiles $\chi_\mu(x)$, $\chi_T(x)$, and $\chi_\star(x)$.
The density profile $\rho(x)$ of the LJ liquid near each wall displays
depletion (negative adsorption) characteristic of solvophobic
substrates \cite{stewart2012pre,
  stewart2014jcp,evans2015jpcm,evans2015prl,
  chacko2017,evans2017,evans2019pnas,rensing2019pnas,evans2016prl},
cf.~the inset in Fig.~\ref{figLJ}(a).  As reported by Evans and
coworkers, $\chimu(x)$ is indeed a better indicator for the emergence
of drying than $\rho(x)$ is.  To see this, consider that the amplitude
of the signal, i.e.\ the enhancement of local fluctuations over the
bulk that is apparent in $\chi_\mu$, but also in $\chi_T$ and in
$\rhostar$ (main plot of Fig.~\ref{figLJ}(a)) is quantitatively much
stronger than the small depletion effect that occurs in the local
density near the wall (inset in Fig.~\ref{figLJ}(a)).  This
sensitivity is intimately linked to the thermodynamic derivative
structure
\eqref{EQparametricDerivative1}--\eqref{EQparametricDerivative3},
which, in a Taylor series sense, probes the local environment around
statepoint $T,\mu$. Clearly, this provides a mechanism to sense the
proximity of a phase transition.  We have checked that both routes to
the fluctuation profiles, via thermodynamic differentiation according
to \eqref{EQparametricDerivative1} and
\eqref{EQparametricDerivative2}, as well as covariance sampling
according to \eqref{EQchimuAsCorrelator} and
\eqref{EQchiTAsCorrelatorEnergyForm} give identical results within
numerical accuracy.  In Fig.~\ref{figLJ}, each fluctuation profile
$\chi_\alpha(x)$ is normalized by its respective bulk value,
$\chi_\alpha^b$, which we obtain consistently from either an
independent bulk simulation run (without external potential and with
periodic boundaries in all three spatial directions) or from the
plateau value at the center of the slit.

Remarkably, $\chiT(x)$ also displays a very strong response near each
wall; recall that this quantity is indicative of entropic correlation
effects, cf.\ \eqref{EQchiTAsCorrelatorEntropicForm}.  Notably, the
entropic fluctuations are much increased when the full temperature
dependence is taken into account, cf.\ the stronger signal of
$\chi_T(x)$ when using $\Lambda(T)$ \cite{SMchilocal} (light orange
line), as compared to $\Lambda=\sigma$ (dark orange line). The reduced
density $\rhostar(x)$ acquires much oscillatory behaviour (see inset
in Fig.~\ref{figLJ}(a)) and it possesses a form that is markedly
different from the density profile~$\rho(x)$, in the present case of
the LJ system.  Note that either convention for $\Lambda$ carries the
full information. Upon changing the convention for $\Lambda$ from
$\sigma \to \Lambda(T)$, the fluctuation profiles acquire kinetic
terms according to $\chi_T(\rv) \to \chi_T(\rv) + k_BD\chimu(\rv)/2$
and $\rhostar(\rv) \to \rhostar(\rv) - k_BTD\chimu(\rv)/2$, with
$\rho(\rv)$ and $\chi_\mu(\rv)$ remaining unchanged. We adhere to
$\Lambda=\sigma$ in the following (which implies considering only
configurational contributions in
\eqref{EQchiTAsCorrelatorEntropicForm}--\eqref{EQrhoStarAsCorrelator}).

The fluctuation profiles at a (quasi-)free gas-liquid interface in the
LJ system are shown in Fig.~\ref{figLJ}(b). In order to stabilize
gas-liquid coexistence in our grand canonical simulation setup, we use
a weak and slowly oscillating external potential, $V_{\rm ext}(x) =
0.2 \epsilon \cos \left( 2\pi x/L_x \right)$, such that the local
chemical potential $\tilde\mu-V_{\mathrm{ext}}(x)$ crosses over from
the gas to the liquid side of the gas-liquid binodal. We use a
periodic simulation box that is extendend in the $x$-direction to $L_x
= 25\sigma$ and choose $\tilde\mu = -3.2\epsilon$ and $k_BT =
0.85\epsilon$. All three $\chi_\alpha(x)$ display a marked signal at
the interface.

Returning to the asymmetric slit pore, we next consider a confined
fluid of hard spheres (HS) of diameter~$\sigma$. Results for the
statepoint $k_BT = 1\epsilon$ and $\tilde\mu=3.36\epsilon$ are shown
in Fig.\ \ref{figLJ}(c).  At the planar hard wall
\cite{davidchack2016,roth2010review} at $x=15\sigma$, and more
generally when all interactions are of hard core type, the fluctuation
profiles simplify, as both interparticle and external potential energy
vanish for all allowed microstates. Comparing the resulting form of
the thermal susceptibility~\eqref{EQchiTAsCorrelatorEnergyForm} with
the definition of the chemical susceptibility (local compressibility)
\eqref{EQchimuAsCorrelator} yields the (hard core) identity
$T\chiT(\rv)= \mu \chimu(\rv)$.  Furthermore, if all (internal and
external) interactions are hard core, the reduced density
\eqref{EQrhoStarAsCorrelator} remarkably simplifies to
$\rhostar(\rv)=\rho(\rv)$. (Kinetic terms can be regained by
transforming to $\Lambda(T)$ as described above.)  The results shown
in Fig.~\ref{figLJ}(c) confirm these properties and illustrate the
spatial variation of the fluctuation fields both at the hard wall and
the soft LJ wall.  For hard spheres and soft external potentials, it
is straightforward to show from \eqref{EQrhoStarAsCorrelator} that
$\rhostar(\rv)=\rho(\rv)-\beta\int d\rv' V_{\rm
  ext}(\rv')\cov(\hat\rho(\rv),\hat\rho(\rv'))$. Note that the
covariance is a fundamental correlator of density fluctuations
\cite{hansen2013,evans1979}.

To further assess how specific the fluctuation profiles are to the
particular type of model fluid, we consider the Gaussian core model
(GCM) \cite{stillinger1976,archer2001,archer2002}, where particles are
allowed to penetrate each other at a finite energy cost. The
interparticle interaction potential has a Gaussian form,
$\phi_{\mathrm{GCM}}(r) = \epsilon {\rm e}^{-r^2/(2 \sigma^2)}$; we
cut off and shift at a distance of $3\sigma$.  Results for the
fluctuation profiles of the GCM are shown in Fig.~\ref{figLJ}(d), for
$\tilde\mu=2.16\epsilon$ and $k_BT = 0.5\epsilon$. The profiles differ
very markedly from those of both the LJ and HS cases. Note in
particular the sign change of $\chi_T(x)$ as compared to the HS case
shown in Fig.~\ref{figLJ}(c).  In summary, on the basis of the
simulation data, we conclude that all three fluctuation profiles are
highly useful quantitative indicators of molecular structuring
phenomena over and beyond the density profile.  Of course,
Fig.~\ref{figLJ}(a) and (b) are most revealing since the density
profile is fairly smooth whereas the other profiles show considerable
structure.

We next turn to addressing the fundamental status of the fluctuation
profiles in more depth. In order to do so, we resort to classical DFT
as the primary modern framework for the predictive description of the
behaviour of inhomogeneous liquids. As a starting point for
constructing functional relations, one often takes the grand
potential, in its elementary Statistical Mechanics form
$\Omega(\mu,V,T)=-k_BT\ln\Xi$, and considers the change due to a
perturbation of the external potential $V_{\rm ext}(\rv)$ at
position~$\rv$. Standard functional calculus demonstrates that the
result is the equilibrium density profile,
\begin{align}
  \rho(\rv) &= \frac{\delta\Omega}{\delta V_{\rm ext}(\rv)}
  \Big|_{\mu VT}.
  \label{EQrhoAsDerivative}
\end{align}
Here $\Omega$ is trivially functionally dependent on $V_{\rm
  ext}(\rv)$ via its occurrence in the Boltzmann factor as the
integrand which yields the partition sum $\Xi$.

The grand potential consists of a sum of energetic, entropic, and
chemical contributions, $\Omega = U-TS-\mu\bar N$, such that $U-TS$ is
the (total) Helmholtz free energy, where $U = \langle \hat H \rangle$
is the average energy and $\bar N = \langle N \rangle$.  Given the
respective definitions of $U,S$ and $\bar N$ in the grand ensemble, it
is straightforward to show that
\begin{align}
  \chimu(\rv) &=  -\frac{\delta \bar N}{\delta V_{\rm ext}(\rv)}
  \Big|_{\mu VT},
  \label{EQchimuAsDerivative}\\
  \chiT(\rv) &=  -\frac{\delta S}{\delta V_{\rm ext}(\rv)}
  \Big|_{\mu VT},
  \label{EQchiTAsDerivative}\\
  \rhostar(\rv)  &= \frac{\delta U}{\delta V_{\rm ext}(\rv)}
  \Big|_{\mu VT},
  \label{EQrhoStarAsDerivative}
\end{align}
which establishes $\chimu(\rv)$ as the response of the total particle
number, $\chiT(\rv)$ as the response of the total entropy, and
$\rhostar(\rv)$ as the response of the total energy upon changing
$V_{\rm ext}(\rv)$ at fixed $\mu,V$ and $T$.  Combining
\eqref{EQchimuAsDerivative}--\eqref{EQrhoStarAsDerivative} and
observing \eqref{EQrhoAsDerivative} and $\Omega=U-TS-\mu\bar N$, it
becomes apparent that the density profile obtained via
\eqref{EQrhoAsDerivative} results from a sum of three distinct
contributions, $\rho(\rv) = \rhostar(\rv) + T\chiT(\rv) +
\mu\chimu(\rv)$, as is consistent
with~\eqref{EQparametricDerivative4}.

In DFT one proceeds by constructing a functional map $\rho(\rv)\to
V_{\rm ext}(\rv)$ which implies that the grand potential is a
functional of the density profile.  The central minimization principle
then yields an Euler-Lagrange equation for the density profile, given
by
\begin{align}
  \ln\big(\Lambda^D\rho(\rv)\big) &= 
  - \beta V_{\rm ext}(\rv) + \beta\mu + c_1(\rv,T),
  \label{EQeulerLagrangeDFT}
\end{align}
where the one-body direct correlation functional is given by the
derivative $c_1(\rv,T)=-\delta \beta F_{\rm
  exc}([\rho],T)/\delta\rho(\rv)$; here the excess (over ideal)
intrinsic free energy functional $F_{\rm exc}(T,[\rho])$ is unique for
a given interparticle interaction potential $u(\rv^N)$. In practical
DFT applications, one chooses an approximation for $F_{\rm exc}[\rho]$
and then solves \eqref{EQeulerLagrangeDFT} for the self-consistent
$\rho(\rv)$ at the given values of $T$ and $\mu$. (Note that $c_1$
depends functionally on $\rho(\rv)$.)

As the Euler-Lagrange equation \eqref{EQeulerLagrangeDFT} holds for
all values of $\mu$ and $T$, with the corresponding equilibrium
density profile $\rho(\rv)$, we can differentiate both sides of
\eqref{EQeulerLagrangeDFT} with respect to either $\mu$ or~$T$.  Via
the functional chain rule (as can be done in nonequilibrium
\cite{pft2013, brader2013noz, brader2014noz}) one obtains two OZ
equations:
\begin{align}
  \chimu^{\rm exc}(\rv) &=
  \rho(\rv) \int d\rv' c_2(\rv,\rv') \chimu(\rv'),
  \label{EQozMu}\\
  \chiT^{\rm exc}(\rv) &=
  \rho(\rv)\Big(\frac{c_1(\rv)}{T} +
  \frac{\partial c_1(\rv)}{\partial T}
  +\int d\rv' c_2(\rv,\rv') \chiT(\rv')\Big),
  \label{EQozT}
\end{align}
where the excess susceptibilities are defined as $\chimu^{\rm
  exc}(\rv)=\chimu(\rv)-\chimu^{\rm id}(\rv)$ and $\chiT^{\rm
  exc}(\rv)=\chiT(\rv)-\chiT^{\rm id}(\rv)$, with the ideal gas
results $\chimu^{\rm id}(\rv)$ and $\chiT^{\rm id}(\rv)$
\cite{SMchilocal}.
The inhomogeneous two-body direct correlation function is given by
$c_2(\rv,\rv')=\delta c_1(\rv)/\delta \rho(\rv') \equiv -\delta^2\beta
F_{\rm exc}[\rho]/\delta\rho(\rv)\delta\rho(\rv')$.  For hard spheres
$\partial c_1^{\rm HS}(\rv)/\partial T=0$, which simplifies
\eqref{EQozT}.  The relation \eqref{EQozMu} generalizes a result for
planar symmetry by Tarazona and Evans \cite{tarazona1982} obtained via
integration over the inhomogeneous OZ equation; their strategy also
leads to the general form \eqref{EQozMu} \cite{coe2020private}.  As
compared to the inhomogeneous OZ relation for the (inhomogeneous) pair
distribution function \cite{hansen2013}, both \eqref{EQozMu} and
\eqref{EQozT} have remarkably simpler, one-body structure. The
striking role of $c_2$ in \eqref{EQozMu} and \eqref{EQozT} as
mediating nonlocal fluctuation effects is consistent with its role in
the inhomogeneous OZ relation.

In summary, we have presented a description of inhomogeneous liquids
based on three one-body fluctuation profiles. In future work, it would
be interesting to relate to the internal-energy functional
\cite{schmidt2011pre}, and to quantum mechanical systems, where the
``softness'' \cite{parr1989} represents a concept similar to
$\chi_\mu(\rv)$. Investigating the role of all $\chi_\alpha(\rv)$ for
drying \cite{coe2020private}, in complex geometries
\cite{giacomello2016,giacomello2019} and in charged systems
\cite{limmer2013} would be highly interesting, as would be devising
new DFT approximation schemes for local fluctuations, possibly based
on machine learning \cite{lin2019,lin2020} or the recent
Barker-Henderson functional \cite{tschopp2020}.

More specifically, the systematic study of all three fluctuation
profiles might help to elucidate which type of density correlations,
whether particle number \eqref{EQparametricDerivative1}, entropy
\eqref{EQparametricDerivative2}, or energy
\eqref{EQparametricDerivative4}, are relevant for hydrophobicity at
the nanoscale \cite{levesque2012jcp,jeanmairet2013jcp,stewart2012pre,
  stewart2014jcp,evans2015jpcm,evans2015prl,
  chacko2017,evans2017,evans2019pnas,rensing2019pnas,evans2016prl,
  giacomello2016,giacomello2019}; work along these lines is in
progress for drying \cite{coe2020}.  Moreover, whether the observed
enhanced fluctuations are a mere consequence of a local decrease in
density, or rather the increase in fluctuations near a hydrophobic
solute forms the underlying physical mechanism for the density
depletion is an interesting question.  Furthermore the OZ relations
\eqref{EQozMu} and \eqref{EQozT} provide a direct and practical link
between the inter-particle structure, as embodied in $c_2$, and the
behaviour of the fluctuation profiles.  Equations \eqref{EQozMu} and
\eqref{EQozT} constitute both a natural bridge towards inhomogeneous
liquid integral equation theory \cite{hansen2013,brader2008}, but they
also suggest the possibility of a stand-alone one-body fluctuation
framework, possibly flanked by generalized density functionals
\cite{schmidt2011pre,tschopp2020,anero2013} or by the transfer of
established \cite{hansen2013}, as well as the development of new,
closure relations for the one-body level.  Beyond inhomogeneous fluids
\cite{evans2016specialIssue,nygard2016prx, martinjimenez2017natCom,
  hernandez-munoz2019,muscatello2017,
  jeanmairet2013jpcl,levesque2012jcp,
  sergiievskyi2014,jeanmairet2019capacitor}, the fluctuation profiles
are uncharted territory in freezing and precursors
\cite{brader2008}. One certainly would expect to find markedly
different behaviour for crystals of hard \cite{haertel2012} and soft
particles \cite{mladek2006}.

\begin{acknowledgments}
Useful discussions and exchanges with Daniel de las Heras, Bob Evans,
Mary Coe, Nigel Wilding, Andrew Archer, Joseph Brader, Roland Roth,
Sophie Hermann, and Thomas \mbox{Fischer} are gratefully acknowledged.
\end{acknowledgments}

\vspace{2mm}
$^*$ Authors contributed equally to this work.

\clearpage

\pagestyle{empty}
\widetext

{\large \bf \noindent 
  Supplemental Material for: Fluctuation profiles in inhomogeneous fluids}\\

{\noindent
  Tobias Eckert, Nico C. X. Stuhlm\"uller, Florian Samm\"uller,
  and Matthias Schmidt}\\

\vspace{-4mm}
{\noindent {\it  Theoretische Physik II, Physikalisches Institut,
  Universit{\"a}t Bayreuth, D-95447 Bayreuth, Germany}\\
  (23 November 2020)
}

\appendix

\section{Ideal Gas}
\label{appendixIdealGas} 

To illustrate the fluctuation profiles, we consider the ideal gas
($u(\rv^N)\equiv 0$) in the presence of an external potential $V_{\rm
  ext}(\rv)$.  The resulting density profile follows a generalized
barometric law \cite{hansen2013},
\begin{align}
  \rho_{\rm id}(\rv) &= \Lambda^{-D}\exp(-\beta(V_{\rm
    ext}(\rv)-\mu)),
\end{align}
where $\Lambda(T)=(2\pi\beta\hbar^2/m)^{1/2}$ indicates the thermal de
Broglie wavelength; here $\hbar$ denotes the reduced Planck constant
and $m$ is the particle mass. One readily obtains the fluctuation
profiles from
\eqref{EQparametricDerivative1}--\eqref{EQparametricDerivative3} [in
  the main text] as
\begin{align}
  \chimu^{\rm id}(\rv)&=\beta\rho(\rv),
  \label{EQchimuIdeal}\\
  \chiT^{\rm id}(\rv)&=
  \frac{\rho(\rv)}{T}
  \big( \beta V_{\rm ext}(\rv)-\beta\mu + D/2\big),
  \label{EQchiTIdeal}\\
  \rhostar^{\rm id}(\rv) &=
  \rho(\rv) - \big(\beta V_{\rm ext}(\rv)+D/2\big)\rho(\rv).
  \label{EQrhoStarIdealrhoForm}
\end{align}
In the context of the OZ relations, \eqref{EQchimuIdeal} and
\eqref{EQchiTIdeal} are relevant beyond the ideal gas, when
$\rho(\rv)$ is taken to be general. For the ideal case,
\begin{align}
  \rho_{\rm id}(\rv) &= \Lambda^{-D}\exp\left(\frac{D}{2}
  -\frac{\chi_T(\rv)}{k_B\chi_\mu(\rv)}\right),
  \label{EQrhoIdOfChis}
  \\
  \rhostar^{\rm id}(\rv)  &= \frac{1-D/2}{\Lambda^D}
  \exp\left(
  \frac{D}{2} 
  -\frac{\chiT(\rv)}{k_B\chimu(\rv)}
  \right)  - V_{\rm ext} \chimu(\rv),
  \label{EQrhoStarIdeal}
\end{align}
where \eqref{EQrhoStarIdeal} follows from
\eqref{EQrhoStarIdealrhoForm} upon using \eqref{EQchimuIdeal},
\eqref{EQchiTIdeal} and \eqref{EQrhoIdOfChis} for $D\neq 2$.  The
Legendre transform~\eqref{EQparametricDerivative3} is hence complete,
including the replacement of the original variables $T,\mu$ by the new
variables $\chiT,\chimu$. For completeness, the temperature of the
inhomogeneous ideal gas can be expressed for $D\neq 2$ via the local
susceptibilities as
\begin{align}
  T &= (k_B\chi_\mu)^{2/(D-2)}
  \left(\frac{mk_B}{2\pi\hbar^2}\right)^{D/(2-D)}
  \exp\left(\frac{2}{2-D}
  \Big(\frac{D}{2}-\frac{\chi_T}{k_B\chi_\mu}\Big)\right),
\end{align}
where position arguments of $\chimu(\rv)$ and $\chiT(\rv)$ are omitted
for clarity.  The thermal wavelength satisfies
\begin{align}
  \Lambda^{D-2} =
  \frac{m}{2\pi\hbar^2 \chi_{\mu}} \exp \left( \frac{D}{2} -
  \frac{\chi_T}{k_B \chi_\mu} \right).
\end{align}  

Throughout, we have retained the temperature dependence of
$\Lambda(T)$. Corresponding results for the fluctuation profiles
within the frequently used convention of measuring lengths against a
molecular size $\sigma$ as the fundamental scale and fixing
$\Lambda=\sigma$ from the outset are obtained by formally setting
$D=0$ in \eqref{EQchiTIdeal}--\eqref{EQrhoStarIdeal}.


\begin{thebibliography}{31}

\bibitem{hansen2013}
  J. P. Hansen and I. R. McDonald, 
  {\it Theory of Simple Liquids}, 4th ed. (Academic Press, London, 2013).

\bibitem{evans1979}
  The nature of the liquid-vapour interface and other topics in
  the statistical mechanics of non-uniform, classical fluids.
  R. Evans, Adv. Phys. \textbf{28}, 143 (1979).
 
\bibitem{evans2016specialIssue}
  New developments in classical density functional theory.
  R. Evans, M. Oettel, R. Roth, and G. Kahl, 
  J. Phys.: Condens. Matter {\bf 28}, 240401 (2016).

\bibitem{nygard2016prx}
  Density Fluctuations of Hard-Sphere Fluids in Narrow Confinement.
  K. Nygard, S. Sarman, K. Hyltegren, S. Chodankar, E. Perret, 
  J. Buitenhuis, J. F. van der Veen, and R. Kjellander,
  Phys. Rev. X {\bf 6}, 011014 (2016).

\bibitem{martinjimenez2017natCom} 
  Atomically resolved three-dimensional structures of
  electrolyte aqueous solutions near a solid surface.
  D. Martin-Jimenez, E. Chac\'on, P. Tarazona, and R. Garcia,
  Nat. Comms. {\bf 7}, 12164 (2016).

\bibitem{hernandez-munoz2019}
  Density functional analysis of atomic force microscopy in a dense fluid.
  J. Hern\'andez-Mu\~noz, E. Chac\'on, and P. Tarazona,
  J. Chem. Phys. {\bf 151}, 034701 (2019).

\bibitem{muscatello2017}
  Deconstructing Temperature Gradients across Fluid
  Interfaces: The Structural Origin of the Thermal Resistance of
  Liquid-Vapor Interfaces.
  J. Muscatello, E. Chac\'on, P. Tarazona, and F. Bresme,
  Phys. Rev. Lett. {\bf 119}, 045901 (2017).

\bibitem{paliwal2018}
  Chemical potential in active systems: predicting phase equilibrium from 
  bulk equations of state?
  S. Paliwal, J. Rodenburg, R. van Roij, and M. Dijkstra,
  New J. Phys. {\bf 20}, 015003 (2018).

\bibitem{rodenburg2018}
  Ratchet-induced variations in bulk states of an active ideal gas.
  J. Rodenburg, S. Paliwal, M. de Jager, P. G. Bolhuis, M. Dijkstra,
  and R. van Roij,
  J. Chem. Phys. {\bf 149}, 174910 (2018).

\bibitem{brader2011shear}
  Density profiles of a colloidal liquid at a wall under shear flow.
  J. M. Brader and M. Krueger,
  Mol. Phys. {\bf 109}, 1029 (2011).

\bibitem{scacchi2018}
 Local phase transitions in driven colloidal suspensions.
 A. Scacchi and J. M. Brader,
 Mol. Phys. {\bf 116}, 378 (2018).

\bibitem{jeanmairet2013jpcl}
  Molecular Density Functional Theory of Water.
  G. Jeanmairet, M. Levesque, R. Vuilleumier, and D. Borgis,
  J. Phys. Chem. Lett. {\bf 4}, 619 (2013).

\bibitem{sergiievskyi2014}
  Fast Computation of Solvation Free Energies with Molecular 
  Density Functional Theory: Thermodynamic-Ensemble Partial 
  Molar Volume Corrections.
  V. P. Sergiievskyi, G. Jeanmairet, M. Levesque, and D. Borgis,
  J. Phys. Chem. Lett. {\bf 5}, 1935 (2014).
  
\bibitem{jeanmairet2019capacitor}
  Study of a water-graphene capacitor with molecular density functional theory.
  G. Jeanmairet, B. Rotenberg, D. Borgis, and M. Salanne,
  J. Chem. Phys. {\bf 151}, 124111 (2019).

\bibitem{levesque2012jcp}
  Scalar fundamental measure theory for hard spheres in three
  dimensions: Application to hydrophobic solvation.
  M. Levesque, R. Vuilleumier, and D. Borgis,
  J. Chem. Phys. {\bf 137}, 034115 (2012).

\bibitem{jeanmairet2013jcp}
  Molecular density functional theory of water describing 
  hydrophobicity at short and long length scales.
  G. Jeanmairet, M. Levesque, and D. Borgis,
  J. Chem. Phys. {\bf 139}, 154101 (2013).

\bibitem{evans2015jpcm}
  The local compressibility of liquids near non-adsorbing substrates: a
  useful measure of solvophobicity and hydrophobicity?
  R. Evans and M. C. Stewart, J. Phys.: Condens. Matter {\bf 27}, 194111
  (2015).

\bibitem{evans2015prl}
  Quantifying Density Fluctuations in Water at a Hydrophobic Surface:
  Evidence for Critical Drying.
  R. Evans and N. B. Wilding, Phys. Rev. Lett. {\bf 115}, 016103 (2015).

\bibitem{chacko2017}
  Solvent fluctuations around solvophobic, solvophilic, and patchy
  nanostructures and the accompanying solvent mediated interactions.
  B. Chacko, R. Evans, and A. J. Archer,
  J. Chem. Phys. {\bf 146}, 124703 (2017).

\bibitem{evans2017}
  Drying and wetting transitions of a Lennard-Jones fluid: Simulations
  and density functional theory.
  R. Evans , M. C. Stewart, and N. B. Wilding,
  J. Chem. Phys. {\bf 147}, 044701 (2017).

\bibitem{evans2019pnas}
  A unified description of hydrophilic and superhydrophobic surfaces in terms of 
  the wetting and drying transitions of liquids.
  R. Evans, M. C. Stewart, and N. B. Wilding,
  Proc. Nat. Acad. Sci. {\bf 116}, 23901 (2019).

\bibitem{rensing2019pnas}
  Commentary: Playing the long game wins the cohesion-adhesion rivalry.
  R. C. Remsing,
  Proc. Nat. Acad. Sci.  {\bf 116}, 23874 (2019).

\bibitem{evans2016prl}
  Critical Drying of Liquids.
  R. Evans, M. C. Stewart, and N. B. Wilding,
  Phys. Rev. Lett. {\bf 117}, 176102 (2016).

\bibitem{stewart2012pre}
  Phase behavior and structure of a fluid confined between competing
  (solvophobic and solvophilic) walls.
  M. C. Stewart and R. Evans,
  Phys. Rev. E {\bf 86}, 031601 (2012).

\bibitem{stewart2014jcp}
  Layering transitions and solvation forces in an asymmetrically confined fluid.
  M. C. Stewart and R. Evans,
  J. Chem. Phys. {\bf 140}, 134704 (2014).

\bibitem{giacomello2016}
  Perpetual superhydrophobicity.
  A. Giacomello, L. Schimmele, S. Dietrich, and M. Tasinkevych,
  Soft Matter {\bf 12}, 8927 (2016).

\bibitem{giacomello2019}
  Recovering superhydrophobicity in nanoscale and macroscale surface textures.
  A. Giacomello, L. Schimmele, S. Dietrich, and M. Tasinkevych,
  Soft Matter {\bf 15}, 7462 (2019).

\bibitem{ornstein1914}
  L. S. Ornstein and F. Zernike,
  Proc. Acad. Sci. Amsterdam {\bf 17}, 793 (1914);
  this article is reprinted in H. Frisch and J. L. Lebowitz, 
  The Equilibrium Theory of Classical Fluids (Benjamin, New York, 1964).

\bibitem{schmidt2011pre}
  Statics and dynamics of inhomogeneous liquids via the 
  internal-energy functional.
  M. Schmidt, Phys. Rev. E {\bf 84}, 051203 (2011). 

\bibitem{SMchilocal} See Supplemental Material, Appendix
  \ref{appendixIdealGas} for the ideal gas fluctuation profiles.

\bibitem{davidchack2016}
  Hard spheres at a planar hard wall: Simulations and density functional theory.
  R. L. Davidchack, B. B. Laird, and R. Roth,
  Condens. Matt. Phys. {\bf 19}, 23001 (2016).

\bibitem{roth2010review}
  Fundamental measure theory for hard-sphere mixtures: a review.
  R. Roth, J. Phys.: Condens. Matt. {\bf 22}, 063102 (2010).

\bibitem{stillinger1976}
  Phase transitions in the Gaussian core system.
  F. H. Stillinger, J. Chem. Phys. {\bf 65}, 3968 (1976).

\bibitem{archer2001}
  Binary Gaussian core model: Fluid-fluid phase separation
  and interfacial properties.
  A. J. Archer and R. Evans,
  Phys. Rev. E {\bf 64}, 041501 (2001).

\bibitem{archer2002}  
  Microscopic theory of solvent-mediated long-range forces: Influence of wetting.
  A. J. Archer, R. Evans, and R. Roth,
  EPL {\bf 59}, 526 (2002).

\bibitem{pft2013} 
  Power functional theory for Brownian dynamics.
  M. Schmidt and J. M. Brader, J. Chem. Phys. {\bf 138}, 214101 (2013).

\bibitem{brader2013noz}
  Nonequilibrium Ornstein-Zernike relation for Brownian many-body dynamics.
  J. M. Brader and M. Schmidt, J. Chem. Phys. {\bf 139}, 104108 (2013). 

\bibitem{brader2014noz}
  Dynamic correlations in Brownian many-body systems.
  J. M. Brader and M. Schmidt, J. Chem. Phys. {\bf 140}, 034104 (2014).

\bibitem{tarazona1982}
  Long ranged correlations at a solid-fluid interface A signature of the
  approach to complete wetting.
  P. Tarazona and R. Evans, Mol. Phys. {\bf  47}, 1033 (1982);
  see their Eqs.~(33) and (39).

\bibitem{parr1989}
  R. G. Parr and W. Yang
  {\it Density-Functional Theory of Atoms and Molecules}
  (Oxford University Press, New York, 1989).

\bibitem{coe2020private}
  M. Coe and R. Evans, private communication.

\bibitem{limmer2013}
  Charge fluctuations in nanoscale capacitors.
  D. T. Limmer, C. Merlet, M. Salanne, D. Chandler, 
  P. A. Madden, R. van Roij, and B. Rotenberg,
  Phys. Rev. Lett. {\bf 111}, 106102 (2013).

\bibitem{lin2019}
  A classical density functional from machine learning and a
  convolutional neural network.
  S.-C. Lin and M. Oettel,
  Scipost Phys. {\bf 6}, 025 (2019).

\bibitem{lin2020}
  Analytical classical density functionals from an equation learning network.
  S.-C. Lin, G. Martius, and M. Oettel,
  J. Chem. Phys. {\bf 152}, 021102 (2020).

\bibitem{tschopp2020}
  Mean-Field Theory of Inhomogeneous Fluids.
  S.~M. Tschopp, H.~D. Vuijk, A. Sharma, and J.~M. Brader,
  Phys. Rev. E {\bf 102}, 042140 (2020).

\bibitem{coe2020}
  M. K. Coe, R. Evans, and N. B. Wilding
  (to be published).

\bibitem{brader2008}
  Structural precursor to freezing: An integral equation study.
  J. M. Brader, J. Chem. Phys. {\bf 128}, 104503 (2008).

\bibitem{anero2013}
  Functional thermo-dynamics: A generalization of dynamic 
  density functional theory to non-isothermal situations.
  J. G. Anero, P. Espa\~nol, and P. Tarazona,
  J. Chem. Phys. {\bf 139}, 034106 (2013).

\bibitem{haertel2012}
  Tension and stiffness of the hard sphere crystal-fluid interface.
  A. Haertel, M. Oettel, R. E. Rozas,  S. U. Egelhaaf, J. Horbach, 
  and H. L\"owen,
  Phys. Rev. Lett. {\bf 108}, 226101 (2012).

\bibitem{mladek2006} Formation of polymorphic cluster phases for a class of
  models of purely repulsive soft spheres.
  B. M. Mladek, D. Gottwald, G. Kahl, M. Neumann, and C. N. Likos,
  Phys. Rev. Lett. {\bf 96}, 045701 (2006).

\end{thebibliography}
\end{document}